\documentclass[12pt]{amsart}
\usepackage{amsbsy,amssymb,amsmath,amsthm,amscd,amsfonts,latexsym,amstext,delarray,
amsmath,graphicx} 
\usepackage[margin=1in]{geometry}
\usepackage{color}
\input xypic

\usepackage[all]{xy}
\usepackage{setspace}

\newtheorem{thm}{Theorem}[section]
\newtheorem{prop}[thm]{Proposition}
\newtheorem{cor}[thm]{Corollary}
\newtheorem{lem}[thm]{Lemma}

\newtheorem{defn}[thm]{Definition}
\newtheorem{rem}[thm]{Remark}

\numberwithin{equation}{section}

\def\F{{\mathbb F}}
\def\Q{{\mathbb Q}}
\def\Z{{\mathbb Z}}

\def\R{{\mathbb R}}

\def\P{{\mathbb P}}
\def\A{{\mathbb A}}

\def\cF{{\mathcal F}}

\def\cV{{\mathcal V}}

\def\bG{{\mathbb G}}

\def\bL{{\mathbb L}}

\def\bT{{\mathbb T}}

\def\bV{{\mathbb V}}

\def\G{{\Gamma}}
\def\Conf{{\rm Conf}}

\title{Potts models with magnetic field: arithmetic, geometry, and computation}
\author{Shival Dasu and Matilde Marcolli}
\address{Mathematics Department, Caltech, 1200 E. California Blvd. Pasadena, CA 91125, USA}
\email{sdasu@caltech.edu}
\email{matilde@caltech.edu}

\date{}

\begin{document}
\maketitle

\begin{abstract}
We give a sheaf theoretic interpretation of Potts models with external magnetic field,
in terms of constructible sheaves and their Euler characteristics. We show that the
polynomial countability question for the hypersurfaces defined by the vanishing of
the partition function is affected by changes in the magnetic field: elementary examples
suffice to see non-polynomially countable cases that become polynomially countable
after a perturbation of the magnetic field. The same recursive formula for the
Grothendieck classes, under edge-doubling operations, holds as in the case without 
magnetic field, but the closed formulae for specific examples 
like banana graphs differ in the presence of magnetic field. We give examples
of computation of the Euler characteristic with compact support, for the set of real
zeros, and find a similar exponential growth with the size of the graph. This can be
viewed as a measure of topological and algorithmic complexity. We also consider
the computational complexity question for evaluations of the polynomial, and show
both tractable and NP-hard examples, using dynamic programming. 
\end{abstract}

\tableofcontents

\section{Introduction}

Several combinatorial graph polynomials have physical significance, either as
partition functions of statistical mechanical models on graphs (Ising and Potts
models), or as the Kirchhoff and Symanzik polynomials that appear in the
parametric form of Feynman integrals in perturbative quantum field theory.
In both cases, it is interesting to consider various questions related to the
properties of these polynomials and of the hypersurfaces they define. For
a survey of the quantum field theory case, we refer the reader to \cite{Mar},
and for the case of Potts models, to \cite{AluMa}, \cite{MaSu} and to the
general survey \cite{Sokal}. 

\smallskip

In this paper, we focus on another such polynomial with physical significance:
the \textbf{V}-polynomial, which gives the partition function of the Potts models
with external magnetic field.

\smallskip

After recalling some general facts about these polynomials, we show in \S \ref{SheafSec} that
they admit a sheaf theoretic interpretation as the Euler characteristics of a constructible complex
$\cF^\bullet_\G$ over the graph configuration space $\Conf_\G(X)$ of a smooth projective variety.
This addresses a question posed to the second author by Spencer Bloch.

\smallskip

In \S  \ref{PolyCountSec}, we consider the hypersurfaces defined by the vanishing of the
\textbf{V}-polynomial, and the question of whether these varieties are polynomially
countable, that is, whether the counting of points over finite fields $\F_p$ is a polynomial in $p$.
In the case of quantum field theory, the analogous 
polynomial countability question has drawn a lot of attention in recent years, in
relation to questions on the occurrence of motives and periods in Feynman integrals.
Counterexamples to polynomial countability for the Kirchhoff polynomials of quantum
field theory are very elusive, and only occur for combinatorially complicated graphs 
with a large number of edges and loops (see the recently found examples in 
\cite{Dor} and \cite{Schn}). It is much simpler to find non-polynomially countable
examples in the case of the Potts model partition function, see \cite{MaSu}. As
expected, even smaller graphs give rise to non-polynomially countable hypersurfaces
in the case of the \textbf{V}-polynomial. However, a new phenomenon occurs: polynomial
countability depends on the magnetic field, and can occasionally be restored by modifying
the magnetic field. We illustrate these phenomena in the simplest examples.

\smallskip

In \S \ref{GrothSec} we consider the class in the Grothendieck ring of varieties of the
complement of the hypersurface defined by the vanishing of the \textbf{V}-polynomial.
We show that the same recursive formula for edge-doubling, proved in \cite{AluMa} in the case
without magnetic field, continues to hold in this case. However, the presence of magnetic field
alters the initial terms of the recursion. We compute the resulting closed form of the class for
the case of banana graphs and compare it with the case without magnetic field. As in
\cite{AluMa}, we then focus on the set of real zeros, and its Euler characteristic with
compact support, as a measure of complexity (topological and algorithmic) of the analytic
set of real zeros. We provide simple examples where this quantity grows exponentially with
the size of the graph.

\smallskip

In \S \ref{CompuSec} we consider a different kind of complexity question regarding the
\textbf{V}-polynomials, namely the computational complexity of evaluating at a point. 
Using dynamic programming, we show that line and polygon graphs are tractable, while
full binary trees, and trees that limit to the line are NP-hard.

\medskip
\subsection{The \textbf{V}-polynomial}

The correspondence between the Tutte polynomial and the partition function assumes a zero-field Hamiltonian \cite{ElMoMoff}, which excludes several important cases, including the presence of an external magnetic field.  However, there exists a combinatorial polynomial that is the evaluation of the Potts model with an external field, the \textbf{V}-polynomial. In this paper, we will study the algebraic, topological, and computational complexity of the \textbf{V}-polynomial.

\smallskip

Let $\Gamma$ be a finite graph, with edge set $E(\Gamma)$ and vertex set $V(\Gamma)$.
A vertex weight on $\Gamma$ is a function $\omega: V(\Gamma)\to S$, with $S$ a 
torsion-free abelian semigroup.

We recall from \cite{ElMoMoff} the definition of the \textbf{V}-polynomial. It is a polynomial in
$\Z[ t=(t_e)_{e\in E(\Gamma)}, x=(x_s)_{s\in S}]$, where the $t_e$ are edge variables (edge weights),
and the $x_s$ account for the presence of the magnetic field. We view the \textbf{V}-polynomial as a map 
${\bf V}_\Gamma : \A^{\# E(\Gamma)}\times \A^{\# S} \to \A$.
For a subset $A \subseteq E(\Gamma)$, we denote by $\Gamma_A\subset \Gamma$ 
the subgraph of $\Gamma$ with $V(\Gamma_A)=V(\Gamma)$
and $E(\Gamma_A)=A$. Let $\Gamma_{A,j}$, for $j=1,\ldots, b_0(\Gamma_A)$
be the connected components of $\Gamma_A$. Then the \textbf{V}-polynomial is
defined as 
\begin{equation}\label{Vpoly}
{\bf V}_\Gamma (t,x)= \sum_{A \subseteq E(\Gamma)} \prod_{j=1}^{b_0(\Gamma_A)} x_{s_j} 
\prod_{e\in A} t_e,
\end{equation}
where $s_j=\sum_{v \in \Gamma_{A,j}} \omega(v)$ is the sum of the weights attached to
all the vertices in the $j$-th component.

\smallskip

The \textbf{V}-polynomial is determined recursively by the deletion--contraction relation
\begin{itemize}
\item For an edge $e\in E(\Gamma)$ that is not  a looping edge,
\begin{equation}\label{delconnoloop}
{\bf V}_\Gamma (t,x)={\bf V}_{\Gamma\smallsetminus e} (\hat t,x) + t_e\, {\bf V}_{\Gamma/e}(\hat t,x),
\end{equation} 
with $\hat t$ the vector of edge variables with $t_e$ removed.
\item For a looping edge $e$
\begin{equation}\label{delconloop}
{\bf V}_\Gamma (t,x)=(t_e +1) {\bf V}_{\Gamma\smallsetminus e} (\hat t,x) ,
\end{equation} 
\item If $\Gamma$ consists of a set of vertices $V(\Gamma)$ and no edges, 
$E(\Gamma)=\emptyset$, then
\begin{equation}\label{onlyvert}
{\bf V}_\Gamma(t,x)= \prod_{v\in V(\Gamma)} x_{\omega(v)},
\end{equation}
where $\omega: V(\Gamma)\to S$ is the vertex weight.
\end{itemize}
Relations between the \textbf{V}-polynomial, the $W$-polynomial of \cite{NobWel},
and the multivariable Tutte polynomial are described in \cite{ElMoMoff}.

\medskip
\subsection{The \textbf{V}-polynomial and magnetic field}

The physical interpretation of the \textbf{V}-polynomial as partition function of the
Potts model with magnetic field comes from rewriting the 
Fourtuin-Kastelyn representation \eqref{Vpoly} of the polynomial as the partition function
$$ Z_\G = \sum_{A\subseteq E(\G)} \prod_{j=1}^{b_0(\G_A)} X_{M_{c_j}}  \prod_{a\in A} (e^{-\beta J_e} -1), $$
where $\beta$ is a thermodynamic inverse temperature parameter, the $J_e$ are the nearest-neighbor interaction energies along the edges, and $M$ is the magnetic field vector, with
$$ X_{M_{c_j}} = \sum_{v \in V(\G_{A,j})} e^{-\beta M_v}, $$ 
with $\G_{A,j}$ the $j$-th connected component of the graph $\G_A$.

\section{A sheaf-theoretic interpretation of Potts models with magnetic fields}\label{SheafSec}

It is known (see \cite{EastHugg}) that the values at integers $n$ of the 
chromatic polynomial of a graph are the Euler characteristics of
the graph configuration spaces $\Conf_\G(\P^{n-1})$.

\smallskip

A natural question then arises: 
whether one can similarly interpret the \textbf{V}-polynomial ${\bf V}_\Gamma (t,x)$ of
the Potts model with magnetic field as an
Euler characteristic of some constructible sheaf (or constructible complex) $\cF^\bullet_\G$
on the configuration spaces $\Conf_\G(X)$ of some smooth projective variety $X$.
In this section we give a positive answer to this question. For generalities about
constructible sheaves, we refer the reader to the monographs \cite{Dim2} and \cite{KaSha}.

\smallskip

The graph configuration spaces are defined as follows. Let $X$ be a smooth projective variety
and let $\G$ be a finite graph with no looping edges and no parallel edges. Then one defines
\begin{equation}\label{ConfG}
\Conf_\G(X) = X^{\# V(\G)} \smallsetminus \cup_{e\in E(\G)} \Delta_e,
\end{equation}
where $\Delta_e \subset X^{\# V(\G)}$ is the diagonal $\Delta_e=\{ (x_v)_{v\in V(\G)} \,|\,
x_{s(e)}=x_{t(e)} \}$, where $\{ s(e), t(e) \}=\partial e$. Graph configuration spaces and
their compactifications were recently used in Feynman integral computations in 
relation to periods and motives, \cite{CeMa1}, \cite{CeMa2}.

\smallskip

In order to recover the ${\bf V}_\Gamma (t,x)$ as Euler characteristic $\chi(\Conf_\G(X), \cF^\bullet_\G)$,
it suffices to show that the latter satisfies the deletion-contraction relation \eqref{delconnoloop} and
the normalization \eqref{onlyvert}, which completely characterize the \textbf{V}-polynomial.

\smallskip

\begin{thm}\label{thmVchi}
Let $\G$ be a finite graph with no looping edges and no parallel edges.
Let $S$ be the set of vertex weights. Let $X$ be a smooth projective variety, and, 
for each $s\in S$, let $\cF_s^\bullet$ be a constructible complex on $X$ with 
\begin{equation}\label{xchi}
x_s:=\chi(X,\cF^\bullet_s).
\end{equation}
Assume all the $\chi(X,\cF^\bullet_s)\neq 0$, for all $s\in S$.
For $\Delta_e \subset X_{s(e)} \times X_{t(e)}$ the edge diagonal, and for $\cF_{\omega(s(e))}^\bullet$, $\cF_{\omega(t(e))}^\bullet$ constructible complexes on $X$, let
\begin{equation}\label{tchi}
t_e := - \frac{\chi(\Delta_e, \iota^{!}_{\Delta_e} (\cF_{\omega(s(e))}^\bullet \overset{L}{\boxtimes} \cF_{\omega(t(e))}^\bullet))}
{\chi(X,\cF_{\omega(s(e))+\omega(t(e))}^\bullet)}.
\end{equation}
There is a constructible complex $\cF^\bullet_\G$ on $\Conf_\G(X)$, determined by the
constructible complexes $\cF_s^\bullet$ on $X$, with $s\in S$, such that the \textbf{V}-polynomial satisfies
\begin{equation}\label{Vchi}
{\bf V}_\Gamma (t,x) = \chi(\Conf_\G(X), \cF^\bullet_\G).
\end{equation}
\end{thm}

\proof The hypothesis that the graph $\G$ has no parallel edges and no looping edges
is motivated by the observation that, if $\G$ has at least one looping edge $e$ with
$x_{s(e)}=x_{t(e)}$ then the diagonal associated to that edge is the whole space $\Delta_e=X^{\# V_\Gamma}$
and the configuration space is empty $\Conf_\G(X) \subset X^{\# V(\G)} \smallsetminus \Delta_e =\emptyset$,
hence in such cases the statement is vacuous. Moreover, if $\G$ has multiple parallel edges, say $e_1,e_2$
with $\partial(e_1)=\partial(e_2)$, then $\Delta_{e_1}=\Delta_{e_2}$, and the configuration space
$\Conf_\G(X)$ is the same as that of the same graph where all but one in each set of parallel
edges are removed. Thus, we can assume to work only with graphs with neither parallel nor looping edges.
In the procedure that follows we will perform repeated operations of deleting and contracting edges from
the initial graph. If $\G$ has no parallel edges then all the contractions $\G/e$ have no looping edges.
If some contraction has parallel edges, we interpreted the resulting configuration space equivalently
as the configuration of the corresponding graph with only simple edges, as above, so that looping
edges do not appear when performing further contractions, hence all the resulting configuration 
spaces are non-empty. 

Recall that a continuous function $f$ determines an exceptional inverse image functor $f^{!}$, which
is the right adjoint of $R f_{!}$ (Verdier duality), see \S 3.2 of \cite{Dim2}. In the particular case of
the inclusion $\iota: Z\hookrightarrow X$ of a locally closed subspace, a more explicit description of
the exceptional inverse image functor $\iota^{!}$ is given in Proposition 3.2 of \cite{Dim2} and in
\S 3.1.12 of \cite{KaSha}.  
Let $\iota_{\Delta_e}: \Delta_e \hookrightarrow X_{s(e)}\times X_{t(e)} \hookrightarrow X^{\# V_\G}$
be the inclusion of the diagonal corresponding to an edge $e$, and let $\iota^{!}_{\Delta_e}$ be
the corresponding exceptional inverse image functor.
Let $e$ be an edge in $\G$. Then 
$$ \Conf_{\G\smallsetminus e}(X) = \Conf_\G (X) \cup \Conf_{\G/e}(X) $$
is a Whitney regular stratification (see Example 1.8 of \cite{Dim1}).
Thus, by theorem 4.1.22 of \cite{Dim2}, given a constructible complex $\cF^\bullet$ on
$\Conf_{\G\smallsetminus e}(X)$, we have 
\begin{equation}\label{delconchi}
 \chi(\Conf_\G (X),\cF^\bullet) = \chi(\Conf_{\G\smallsetminus e}(X), \cF^\bullet) 
- \chi(\Conf_{\G/e}(X), \iota^{!}_{\Delta_e} \cF^\bullet). 
\end{equation}
Note that if $e$ is in a set of multiple parallel edges of $\G$ so that
the contraction $\G/e$ has looping edges, the last term in \eqref{delconchi} is 
trivial, since $\Conf_{\G/e}(X)$ is empty, and we simply get the relation 
$\chi(\Conf_\G (X),\cF^\bullet) = \chi(\Conf_{\G\smallsetminus e}(X), \cF^\bullet)$, 
which means we can reduce multiple edges to simple edges, as assumed above.

Iterating this relation by repeatedly removing and contracting edges, until there are no edges left,
we obtain an expression for $\chi(\Conf_\G (X),\cF^\bullet)$ as a sum over
the set of all subsets $A\subseteq E(\G)$, of terms (with signs) 
of the form $\chi(\Conf_{\G_A/A}(X), \iota^{!}_A \cF^\bullet)$, where $\G_A$ is
the graph with vertex set $V(\G_A)=V(\G)$ and edge set $E(\G_A)=A$, while
$\G_A/A$ is the graph obtained from $\G_A$ by contracting all the edges. This
is a graph with $\# V(\G_A/A)=b_0(\G_A)$, the number of connected
components, and with $E(\G_A/A)=\emptyset$. 

Recall that, for a product $X_1 \times X_2$, with $p_i$, $i=1,2$ the projections onto the
two factors, and for constructible complexes $\cF_i^\bullet$ over $X_i$, one denotes by 
$\cF_1^\bullet \overset{L}{\boxtimes} \cF_2^\bullet = p_1^{-1} \cF_1^\bullet \overset{L}{\otimes} 
p_2^{-1} \cF_2^\bullet$ the external left derived tensor product of the $\cF_i^\bullet$, see Corollary
2.3.30 of \cite{Dim2}. We consider on $\Conf_{\G_A/A}(X)=X^{b_0(\G_A)}$
a constructible complex of the form 
$$  \cF^\bullet_{\G_A/A} := \overset{L}{\boxtimes}_{j=1, \ldots, b_0(\G_A)} \cF_{s_j}^\bullet, $$
where the $\cF^\bullet_{s_j}$ are the constructible complexes in the collection $\{ \cF^\bullet_s \}_{s\in S}$, 
where $s_j =\omega(v_j)$ is the weight of the $j$-th vertex $v_j$ of $\G_A/A$. Then, by the K\"unneth
formula of Corollary 2.3.31 of \cite{Dim2}, we have
$$ \chi(\Conf_{\G_A/A}(X), \cF^\bullet_{\G_A/A}) = \prod_{j=1}^{b_0(\G_A)} \chi(X,\cF^\bullet_{s_j}). $$
Thus, we only need then to compare the Euler characteristic
$\chi(\Conf_{\G_A/A}(X), \iota^{!}_A \cF^\bullet)$ with the Euler characteristic 
$\chi(\Conf_{\G_A/A}(X), \cF^\bullet_{\G_A/A})$. 

Recall that, if $v_1$ and $v_2$ are two vertices in a graph $\G$, connected by an edge $e\in E(\G)$
and if $v$ is the corresponding vertex in the contraction graph $\G/e$, then the vertex weights 
satisfy $\omega(v) = \omega(v_1)+\omega(v_2) \in S$, see \S 2.2. of \cite{ElMoMoff}.
Consider the constructible complex $\cF^\bullet = \overset{L}{\boxtimes}_{v\in V(\G)} \cF^\bullet_{\omega(v)}$
on $X^{\# V(\G)}$. We have $\chi(X^{\# V(\G)}, \cF^\bullet)=\prod_v \chi(X,\cF^\bullet_{\omega(v)})$. 
When we contract an edge $e\in A$, we obtain, by definition of the variable $t_e$
$$ \chi(X, \iota^{!}_{\Delta_e} (\cF^\bullet_{\omega(s(e))} \overset{L}{\boxtimes} \cF^\bullet_{\omega(t(e))}) =
- t_e \, \chi(X, \cF^\bullet_{\omega(s(e))+\omega(t(e))}). $$
Let $V(\G_{A,j})$ be the set of vertices in the $j$-th connected component of
$\G_A$.  Inductively, we obtain 
$$ \chi(\Conf_{\G_A/A}(X), \iota^{!}_A \cF^\bullet) = (-1)^{\# A} \prod_{e\in A} t_e  \prod_{j=1}^{b_0(\G_A)} 
\chi(X, \cF^\bullet_{\sum_{v\in V(\G_{A,j})} \omega(v)}). $$
Thus, we obtain 
$$ \chi(\Conf_\G(X), \cF^\bullet) = 
\sum_{A\subseteq E(\G)} \prod_{e\in A} t_e  \prod_{j=1}^{b_0(\G_A)}  x_j . $$
where $x_j= \chi(X, \cF^\bullet_{\sum_{v\in V(\G_{A,j})} \omega(v)})$.
\endproof

\section{Polynomial countability (arithmetic complexity)}\label{PolyCountSec}

In this section we consider the polynomial countability question for the
affine hypersurfaces defined by the vanishing of the \textbf{V}-polynomial.
We show that one can find non-polynomially countable examples already for the 
simplest polygonal graphs. However, we also show in a very simple example
that one can sometime restore polynomial countability by changes to the magnetic field.

\medskip
\subsection{The \textbf{V}-hypersurfaces}

Let $Var(\Gamma)\subset E(\Gamma) \times S$ be the set of indices of variables $(t,x)$ the 
\textbf{V}-polynomial of $\Gamma$ depends on, and let
\begin{equation}\label{Vhyp}
\bV_{\Gamma}=\{ (t,x) \in \A^{\# Var(\Gamma)} \,|\, V_{\Gamma}(t,x)=0 \} \subset 
 \A^{\# Var(\Gamma)}
 \end{equation}
be the affine hypersurface determined by the vanishing of the \textbf{V}-polynomial.
Since the polynomial has integer coefficients (in fact, all coefficients are equal to $1$), we
can regard the variety $\bV_{\Gamma}$ as defined over $\Q$, and in fact over $\Z$.

For the reduction of $\bV_{\Gamma}$ modulo a prime $p$, we let
$N_{\bV_{\Gamma}}(q)$ be the counting function that gives the number
of $\F_q$-points, for a power $q=p^r$. The hypersurface $\bV_{\Gamma}$
is polynomial countable if, for all $p$ and $q=p^r$, the counting function
$N_{\bV_{\Gamma}}(q)$ is a polynomial in $q$ with $\Z$-coefficients.

We can regard the behavior (polynomial or non-polynomial) of the counting 
function $N_{\bV_{\Gamma}}(q)$ as an indicator of the arithmetic (or motivic)
complexity of the hypersurfaces $\bV_{\Gamma}$.

\medskip
\subsection{The polynomial countability question}
For the Kirchhoff polynomial, all graphs with fewer 
than 12 edges were verified to give rise to polynomially countable hypersurfaces in \cite{Stem}, using a method
of \cite{Stan}. However, a general result of \cite{BelBro} showed that the
graph hypersurfaces additively generate a localization of the Grothendieck ring
of varieties, hence most graphs have necessarily non-polynomially countable hypersurfaces.
Indeed, more recently, explicit non-polynomially countable examples (starting
at 14 edges) were obtained in \cite{Dor} and \cite{Schn}.
The analogous polynomial countability question for the hypersurfaces defined by
the Potts model polynomial is considerably simpler. It was shown in \cite{MaSu}
that already the Potts model hypersurface associated to the tetrahedron graph 
$\Gamma=K_4$ is not polynomially countable. Similarly, hypersurfaces defined
by polynomials arising in Ponzano--Regge models of quantum gravity were
shown to be non-polynomially countable for the tetrahedron graph, \cite{DLi}.
Clearly, one expects that polynomial countability will fail even more easily in 
the case of Potts models with
external magnetic fields. This is indeed the case, in the sense that it is very easy to
find very small graphs for which the \textbf{V}-hypersurface is not polynomially countable.
However, a new phenomenon occurs: a perturbation of the magnetic field can
restore polynomial countability.

\subsection{Constant magnetic field and perturbations}

We discuss here the simplest possible example that illustrates the
phenomenon described above. We first consider the triangle graph
with constant external magnetic field. We show that the associated 
\textbf{V}-hypersurface is not polynomially countable. We then show
that changing the magnetic field can restore polynomial
countability.

\smallskip

\begin{lem}\label{Vtriangleconst}
The \textbf{V}-polynomial for the triangle graph with a constant
magnetic field (equal to $1$ at each vertex) is given by
\begin{equation}\label{cVtri}
x_{1}^{3} + (t_{e_{1}} + t_{e_{2}} + t_{e_{3}})x_{2}x_{1} 
+ (t_{e_{1}}t_{e_{2}} + t_{e_{2}}t_{e_{3}} + t_{e_{1}}t_{e_{3}} + t_{e_{1}}t_{e_{2}}t_{e_{3}})x_{3}.
\end{equation}
\end{lem}

\proof This follows directly from \ref{Vpoly}, in the case with a constant magnetic field.
\endproof

\smallskip

\begin{thm}\label{cVtrinonPC}
The \textbf{V}-hypersurface of the triangle graph with a constant magnetic 
field equal to $1$ at each vertex is not polynomially countable.
\end{thm}

\proof We write \eqref{cVtri} equivalently as
\begin{equation}\label{cVtri2}
x_{1}^{3} + (t_{e_{1}} + t_{e_{2}} + t_{e_{3}})x_{2}x_{1} 
+ x_{3}(t_{e_{1}}(t_{e_{2}} + t_{e_{3}} + t_{e_{2}}t_{e_{3}}) + t_{e_{2}}t_{e_{3}}).
\end{equation}
We break the counting of zeros of the polynomial over a finite field $\F_p$ into 
several cases. First, when $x_{1}=0$ then $x_{2}$ can be anything. Still
assuming $x_{1}=0$, the case when $x_{3} = 0$ then contributes $p^{4}$ zeros.  
When $x_{3} \neq 0$ (with $x_{1}=0$), then we have
\begin{equation}
t_{e_{1}}(t_{e_{2}} + t_{e_{3}} + t_{e_{2}} t_{e_{3}}) + t_{e_{2}}t_{e_{3}} = 0.
\end{equation}
If $t_{e_{2}} + t_{e_{3}} + t_{e_{2}}t_{e_{3}} = 0$, then $t_{e_{2}}t_{e_{3}} = 0$, so both $t_{e_{2}}$ and $t_{e_{3}}$ are 0 and $t_{e_{1}}$ can be anything. Thus, this case contributes 
$p^{2}(p-1)$ points. Otherwise, we have $$t_{e_{2}} \neq -\frac{t_{e_{3}}}{t_{e_{3}}+1} \ \  \text{ and } \ \  t_{e_{1}}  = -\frac{t_{e_{2}}t_{e_{3}}}{t_{e_{2}} + t_{e_{3}} + t_{e_{2}}t_{e_{3}}}, $$
contributing $p^{2}(p-1)^{2}$ points.

Now, suppose that $x_{1} \neq 0$. Then, when $t_{e_{1}} + t_{e_{2}} + t_{e_{3}} \neq 0$, we can solve for $x_{2}$ with $x_{3}$ arbitrary. This case contributes $p^{3}(p - 1)^{2}$ points. Now suppose that $t_{e_{1}} = -(t_{e_{2}} + t_{e_{3}})$. 
Then $x_{2}$ can be arbitrary. Substituting for $t_{e_{3}}$, we have 
\begin{equation}\label{subseq}
x_{1}^3 + x_{3}(t_{e_{1}}^2 + t_{e_{2}}^2 + t_{e_{1}}t_{e_{2}} + t_{e_{1}}^2 t_{e_{2}} + t_{e_{1}} t_{e_{2}}^2) = 0.
\end{equation}
Let us set 
\begin{equation}\label{ft1t2}
f(t_{e_{1}},t_{e_{2}}) = t_{e_{1}}^2 + t_{e_{2}}^2 + t_{e_{1}}t_{e_{2}} + t_{e_{1}}^2 t_{e_{2}} + t_{e_{1}} t_{e_{2}}^2. 
\end{equation}
Then, whenever $f(t_{e_{1}}, t_{e_{2}}) \neq 0$, we have that $$x_{3} = \frac{x_{1}^3}{f(t_{e_{1}}, t_{e_{2}})}.$$
This contributes a number of points equal to $(p-1)p(p^2 - Z[f(t_{e_{1}},t_{e_{2}})])$, where 
$Z[f(t_{e_{1}},t_{e_{2}})]$ denotes the number of $\F_p$-points of $f(t_{e_{1}}, t_{e_{2}})=0$.  
Now, arguing as in \cite{Stem}, if $Z[f(t_{e_{1}},t_{e_{2}})]$ were a polynomial, it would have to be linear, 
which it is not by inspection.
\endproof

We now consider the same graph, but with a variable magnetic field. In particular, we look at the case
where the magnetic field is zero at two of the vertices.

\begin{lem}\label{triM01}
The \textbf{V}-polynomial of the triangle graph with a magnetic field of 1 at one of the vertices and 0 
at the others is given by
\begin{equation}\label{eqM01}
x_{1}(x_{0}^2 + x_{0}(t_{e_{1}} + t_{e_{2}} + t_{e_{3}}) + t_{e_{1}}(t_{e_{2}} + t_{e_{3}} + t_{e_{2}} t_{e_{3}}) + t_{e_{2}} t_{e_{3}}).
\end{equation}
\end{lem}

\proof Again, this follows immediately from \ref{Vpoly}.
\endproof

\begin{thm}\label{VtriPC}
The \textbf{V}-polynomial of the triangle graph with a magnetic field of 1 at one of the vertices and 0 
at the others is polynomially countable.
\end{thm}

\proof If $x_{1} = 0$, then the polynomial \eqref{eqM01} is zero, with the other variables free, hence 
contributing $p^{4}$ points over $\F_p$. 
If $x_{1} \neq 0$, then suppose $x_{0} \neq -t_{e_{2}} - t_{e_{3}} - t_{e_{3}}t_{e_{2}}$. In this case we have
$$ t_{e_{1}} = \frac{x_{0}^2 + x_{0}(t_{e_2} + t_{e_3}) + t_{e_{2}}t_{e_{3}}}{x_{0} + t_{e_{2}} + t_{e_{3}} + t_{e_{2}}t_{e_{3}}}, $$ with $t_{e_{2}}$ and  with $t_{e_{3}}$ free. This case thus contributes $(p-1)^{2}p^2$ points.
If $x_{0} = -t_{e_{2}} - t_{e_{3}} - t_{e_{3}} t_{e_{2}}$, then, after substituting, we need to solve for $t_{e_{2}}t_{e_{3}} 
+ t_{e_{2}}t_{e_{3}}^2 + t_{e_{2}}^2 t_{e_{3}} + t_{e_{2}}^2 t_{e_{3}}^2 = 0$. 
Counting the pairs which satisfy this equation, this contributes a total of $(p-1)p(2(p-2) + p)$ zeros. 
Therefore, the total number of zeros of the polynomial is $4 p - 7 p^2 + 2 p^3 + 2 p^4$, 
which is a polynomial in $p$.
\endproof

\bigskip

\section{The Euler characteristic (topological complexity)}\label{GrothSec}

The polynomial countability question considered in the previous section is closely related to the
form of the class $[\bV_{\Gamma}]$ of the hypersurface in the Grothendieck ring of varieties.
Recall that, for $R$ a field or ring, the Grothendieck ring $K_0(\cV_R)$ of varieties over $R$
is generated by the isomorphism classes $[X]$ of smooth quasi-projective varieties, with the
inclusion--exclusion relation $[X]=[Y]+[X\smallsetminus Y]$ for $Y\subset X$ a closed embedding
and with the product given by $[X][Y]=[X\times Y]$. The Lefschetz motive $\bL=[\A^1]$ is
the class of the affine line. If the class of a variety is a polynomial $[X]=\sum_k a_k \bL^k$ in the
Lefschetz motive, then $X$ is polynomially countable, and the converse is conditional to
conjectures on motives.

In the case of the graph polynomials of quantum field theory, \cite{AluMa2}, and of the
Potts model without magnetic field, \cite{AluMa}, recursive formulae were obtained for
the Grothendieck classes of the hypersurface complements, under simple operations
performed on the underlying graph, such as doubling or splitting edges. 

In this section we consider the edge-doubling operation and we show that the
same recursion continues to hold in the case with magnetic field, but the initial
step of the recursion is changed by the presence of the magnetic field. We
compute the resulting classes for the banana graphs.

We then compute the Euler characteristic with compact support of the set of
real solutions, again in the case of the banana graphs, where we show it
grows exponentially with the size of the graph. As shown in \cite{AluMa} (see also
\cite{BuKu}, \cite{Dut}, \cite{Yao}), the Euler characteristic with compact support
can be used as an estimate of algorithmic complexity of the analytic set of real solutions.

\medskip
\subsection{Deletion--contraction relation and classes in the Grothendieck ring}

In both the cases of the graph polynomials of quantum field theory and the
Potts model partition function, the key ingredient in order to obtain recursive
formulae for the Grothendieck classes under iterated edge doubling, is a 
formula relating the deletion-contraction property of the polynomial to a
relation between the classes. One does not have an actual deletion-contraction
relation at the level of Grothendieck classes (for example, because if such a 
relation existed, then the classes would always remain in the Tate part $\Z[\bL]$ of the 
Grothendieck ring, which we know is not the case). However, one obtains 
from \eqref{delconnoloop} a relation
\begin{equation}\label{delconGr}
 [ \A^{\# Var(\Gamma)}\smallsetminus \bV_{\Gamma}]
 = \bL [ \A^{\# Var(\Gamma)-1}\smallsetminus (\bV_{\Gamma\smallsetminus e} \cap \bV_{\Gamma/e})]
 - [ \A^{\# Var(\Gamma)-1}\smallsetminus  \bV_{\Gamma\smallsetminus e}].
\end{equation} 
The remaining cases (bridges and looping edges)
are simpler, and give the class $[\A^{\# Var(\Gamma)}\smallsetminus \bV_{\Gamma}]$
as a function of either the class of the deletion or that of the contraction and the Lefschetz motive,
see \cite{AluMa}.

\medskip
\subsection{Edge doubling and parallel edges recursion}

Let $^{e}\Gamma$ denote the graph obtained from a given graph $\Gamma$ by doubling an edge $e$ that is not
a looping edge. We also denote by $^{m}\Gamma$ the graph obtained by adding $m$ edges parallel to $e$ in 
$\Gamma$. Let $f$ denote the new edge resulting from the doubling of an edge $e$. Using the deletion-contraction 
relation \eqref{delconnoloop} for the \textbf{V}-polynomial, we obtain the following expression for the 
\textbf{V}-polynomials of $^{e}\Gamma$:
\begin{equation}\label{Vef}
V_{^{e}\Gamma} = V_{\Gamma \smallsetminus e} + (t_{e} + t_{f} + t_{e}t_{f})V_{\Gamma /e}.
\end{equation}
As in \cite{AluMa}, we set $u_e=1+t_e$ and $u_f=1+t_f$.
With this change of variables, we can write \eqref{Vef} equivalently as
\begin{equation}
V_{^{e}\Gamma} = V_{\Gamma \smallsetminus e} + (u_{e}u_{f} - 1)V_{\Gamma/e}.
\end{equation}

\smallskip

Let $Var(\Gamma)\subset E(\Gamma) \times S$ be the set of indices of variables $(t,x)$ the 
\textbf{V}-polynomial of $\Gamma$ depends on. This set depends not only on $\Gamma$ 
but also on the assignment of the magnetic field.
Let $\bV_{^{e}\Gamma}=\{ (t,x) \in \A^{\# Var(^{e}\Gamma)} \,|\, V_{^{e}\Gamma}(t,x)=0 \}$ denote the hypersurface in 
$ \A^{\# Var(^{e}\Gamma)}$ determined by the vanishing of the \textbf{V}-polynomial.
Let $\{ \bV_{^{e}\Gamma} \}$ denote the class in the Grothendieck ring of the hypersurface complement
\begin{equation}\label{VhypcomplGr}
\{ \bV_{^{e}\Gamma} \} := [ \A^{\# Var(^{e}\Gamma)} \smallsetminus \bV_{^{e}\Gamma} ] \in K_0(\cV_\Z).
\end{equation}

\smallskip

The following result is the analog, in the case with magnetic field, of Theorem 6.1 of \cite{AluMa}
for the Potts model without magnetic field. Notice that the presence of the magnetic field does not
alter the edge-doubling relation, which remains the same as in the original Potts model case.

\begin{thm}\label{edgedoublethm}
Let $^{e}\Gamma$ be the graph obtained by doubling an edge $e$ in $\Gamma$, that is not a 
looping edge. Then the class $\{ \bV_{^{e}\Gamma} \}$ of the hypersurface complement satisfies
\begin{equation}\label{edge2class}
\{ \bV_{^{e}\Gamma} \}= \mathbb{T}\{ \bV_{\Gamma}\} - (\mathbb{T} +1)\{ Z(V_{\Gamma \smallsetminus e} - V_{\Gamma /e}) \},
\end{equation}
where $\bT=[\bG_m]=\bL-1$ is the class of the multiplicative group, with $\bL$ the Lefschetz motive,
and with $Z(V_{\Gamma\smallsetminus e} - V_{\Gamma/e})$ the locus defined by the vanishing of the polynomial
$V_{\Gamma\smallsetminus e} - V_{\Gamma/e}$.
\end{thm}

\proof The argument is analogous to Theorem 6.1 of \cite{AluMa}. 
We break the computation of the Grothendieck class $\{ V_{^{e}\Gamma} \}$ into 
several cases.
If $V_{\Gamma /e} = 0$, then $V_{\Gamma \smallsetminus e} \neq 0$, and $u_{e}$ and $u_{f}$ are free variables.
Thus, this case contributes a term 
\begin{equation}
 (\mathbb{T}+1)^{2}[ \bV_{\Gamma /e}\smallsetminus ( \bV_{\Gamma/e}\cap \bV_{\Gamma \smallsetminus e})].
\end{equation}
If $V_{\Gamma / e} \neq 0$, then we have one of two possibilities. 
The first is that $V_{\Gamma / e} = V_{\Gamma \smallsetminus e}$, in which case $u_{e}u_{f} \neq 0$, contributing 
a class 
\begin{equation}
    \mathbb{T}^{2}[Z(V_{\Gamma \smallsetminus e} - V_{\Gamma /e})\cap (\mathbb{A}^{\# Var(\Gamma)  - 1}\smallsetminus 
    \mathbb{V}_{\Gamma /e})].
\end{equation}
The other possibility is that $V_{\Gamma / e} \neq V_{\Gamma \smallsetminus e}$. In this case $u_{e}u_{f} \neq c$ for some $c \neq 0$. If $u_e u_f = c$, then $u_f \neq 0$ and $u_e = c/u_f$, which gives a $\mathbb{L} - 1$. Therefore, this term 
contributes a class 
$$
 (\mathbb{T}^{2} + \mathbb{T} + 1)[(\mathbb{A}^{\# Var(\Gamma) - 1}\smallsetminus \mathbb{V}_{\Gamma/e})\smallsetminus Z(V_{\Gamma \smallsetminus e} - V_{\Gamma /e})].
$$
Thus, the class of the hypersurface complement of $V_{^{e}\Gamma}$ is given by
$$
 (\mathbb{T}+1)^{2}[\mathbb{V}_{\Gamma/e}\smallsetminus (\mathbb{V}_{\Gamma/e}\cap
 \mathbb{V}_{\Gamma\smallsetminus e})] + \mathbb{T}^{2}[Z(V_{\Gamma \smallsetminus e} - V_{\Gamma/e})\cap (\mathbb{A}^{\# Var(\Gamma) - 1}\smallsetminus \mathbb{V}_{\Gamma/e})] $$
$$ + (\mathbb{T}^{2} + \mathbb{T} + 1)[(\mathbb{A}^{\# Var(\Gamma) - 1}\smallsetminus \mathbb{V}_{\Gamma/e})\smallsetminus Z(V_{\Gamma\smallsetminus e} - V_{\Gamma/e})].
$$
This can be simplified to
$$
 (\mathbb{T}+1)^{2}[\mathbb{V}_{\Gamma/e}\smallsetminus (\mathbb{V}_{\Gamma/e}\cap\mathbb{V}_{\Gamma\smallsetminus e})] - (\mathbb{T}+1)[Z(V_{\Gamma\smallsetminus e} - V_{\Gamma/e})\cap (\mathbb{A}^{\# Var(\Gamma) - 1}\smallsetminus \mathbb{V}_{\Gamma/e})] $$
$$ + (\mathbb{T}^{2} + \mathbb{T} + 1)[(\mathbb{A}^{\# Var(\Gamma) - 1}\smallsetminus \mathbb{V}_{\Gamma/e})] $$
$$ = (\mathbb{T}^{2} +2\mathbb{T} + 1)[\mathbb{V}_{\Gamma/e}] -   (\mathbb{T}+1)^{2}[\mathbb{V}_{\G/e}\cap\mathbb{V}_{\G\smallsetminus e}] - (\mathbb{T}+1)[Z(V_{\G\smallsetminus e} - V_{\G/e})\cap (\mathbb{A}^{\# Var(\G) - 1}\smallsetminus \mathbb{V}_{\G/e})] $$
$$ + (\mathbb{T}^{2} + \mathbb{T} + 1)[(\mathbb{A}^{\# Var(\G) - 1}\smallsetminus \mathbb{V}_{\G/e})] $$
$$ = \mathbb{T}[\mathbb{V}_{\G/e}] -   (\mathbb{T}+1)^{2}[\mathbb{V}_{\G/e}\cap\mathbb{V}_{\G\smallsetminus e}] - (\mathbb{T}+1)[Z(V_{\G\smallsetminus e} - V_{\G/e})\cap (\mathbb{A}^{\# Var(\G) - 1}\smallsetminus \mathbb{V}_{\G/e})] $$
$$ + (\mathbb{T}^{2} + \mathbb{T} + 1)\mathbb{L}^{\# Var(\G) - 1} $$
$$  = (\mathbb{T} + 1)^{2}(\mathbb{L}^{\# Var(\G) - 1} - [\mathbb{V}_{\G/e}\cap\mathbb{V}_{\G\smallsetminus e}]) - \mathbb{T}(\mathbb{L}^{\# Var(\G) - 1} - [\mathbb{V}_{G/e}])  $$
$$ - (\mathbb{T}+1)[Z(V_{\G\smallsetminus e} - V_{\G/e})\cap (\mathbb{A}^{\# Var(\G) - 1}
 \smallsetminus \mathbb{V}_{\G/e})] . $$
 
We can express $\{\mathbb{V}_{\G/e}\cap\mathbb{V}_{\G\smallsetminus e}\}$ using the deletion-contraction relation
\eqref{delconGr}, obtaining
$$
(\mathbb{T}+1)(\{\mathbb{V}_{\G}\} + \{\mathbb{V}_{\G/e}\}) - \mathbb{T}(\{\mathbb{V}_{\G/e}\}) 
- (\mathbb{T}+1)[Z(V_{\G\smallsetminus e} - V_{\G/e})\cap (\mathbb{A}^{\# Var(\G) - 1}\smallsetminus \mathbb{V}_{\G/e})]
$$
$$
= (\mathbb{T}+1)(\{\mathbb{V}_{\G}\}) + \{\mathbb{V}_{\G/e}\}
- (\mathbb{T}+1)[Z(V_{\G\smallsetminus e} - V_{\G/e})\cap (\mathbb{A}^{\# Var(\G) \vert - 1}\smallsetminus 
\mathbb{V}_{\G/e})].
$$
Then, since we have
 $$ (\mathbb{T}+1)[Z(V_{\G\smallsetminus e} - V_{\G/e})\cap (\mathbb{A}^{\# Var(\G) - 1}\smallsetminus \mathbb{V}_{\G/e})] = Z(V_{\G\smallsetminus e} - V_{\G/e})\smallsetminus(\mathbb{V}_{\G/e}\cap\mathbb{V}_{\G\smallsetminus e}), $$
 applying deletion-contraction once more, we get that the class of the complement is equal to \eqref{edge2class}.
\endproof 
 
As in Theorem 6.4 of \cite{AluMa}, we then obtain the recursion relation for multiple parallel edges. Again, this
is unaltered with respect to the case without magnetic field.

\begin{cor}\label{edgedoublethm}
Let $^{m}\Gamma$ be the graph obtained by adding $m\geq 2$ edges parallel to $e$ in $\Gamma$.
The Grothendieck classes $\{\mathbb{V}_{^{m}\Gamma }\}$ of the hypersurface complements 
satisfy the recursion relation
\begin{equation}\label{medgerecur}
\{\mathbb{V}_{^{m}\Gamma}\} = (2 \mathbb{T} + 1)\{\mathbb{V}_{^{m-1}\Gamma}\} - \mathbb{T}(\mathbb{T} + 1)\{\mathbb{V}_{^{m-2}\Gamma}\}.
\end{equation}
\end{cor}

\proof As in \cite{AluMa}, the relation \eqref{medgerecur} is obtained from the relation for the class
$\{ \bV_{^{e}\Gamma} \}$ of the graph $^{e}\Gamma$ obtained by doubling the edge $e$.
\endproof

Thus, for the case with magnetic field, we obtain the same generating function as in
Theorem 6.5 of \cite{AluMa} for the original case without magnetic field.

\begin{cor}\label{genfunc}
The exponential generating function for the hypersurface complements $\{\mathbb{V}_{^{m}\G}\}$ is
\begin{equation}\label{gen}
\sum\limits_{m\geq 0} \{\mathbb{V}_{^{m}\G}\}\frac{s^{m}}{m!} = ((\mathbb{T} + 1)\{\mathbb{V}_{\G}\} - \{\mathbb{V}_{^{1}\G}\})e^{\mathbb{T}s} + (\{\mathbb{V}_{^{1}\G}\} - \mathbb{T}\{\mathbb{V}_{\G}\})e^{(\mathbb{T} + 1)s}.
\end{equation}
\end{cor}

\proof The argument is the same as in Theorem 6.5 of \cite{AluMa}.
\endproof

\medskip
\subsection{Potts models with magnetic field on banana graphs}

As in \cite{AluMa}, we can use the recursive formula for multiple edges described above to
compute the Grothendieck classes of the hypersurface complements for the banana graphs.
The $n$-th banana graph is the graph with two vertices of valence $n$ and $n$ parallel edges
connecting them.

\begin{lem}\label{ban0}
$^{0}\G$ is the graph with two vertices and a single non-looping edge between them.
The corresponding Grothendieck class is 
\begin{equation}\label{ban0class}
\{\mathbb{V}_{^{0}\G}\} =  (\mathbb{T} + 1)\mathbb{T}^{2} + (\mathbb{T} + 1)^{2}\mathbb{T}^2.
\end{equation}
\end{lem}

\proof The \textbf{V}-polynomial of $^{0}\G$ is given by
\begin{equation}
V_{^{0}\G} = t_{e}x_{c1+c2} + x_{c1}x_{c2}.
\end{equation}
If $x_{c1+c2} = 0$, then $t_{e}$ is free and $x_{c1} x_{c2} \neq 0$, contributing a class 
\begin{equation}
\mathbb{L}(\mathbb{L} - 1)^{2} = (\mathbb{T} + 1)\mathbb{T}^{2}
\end{equation}
If $x_{c1+c2} \neq 0$, then $t_{e} \neq -\frac{x_{c1}x_{c2}}{x_{c1+c2}}$ contributing a class
\begin{equation}
\mathbb{L}^{2}(\mathbb{L} - 1)^{2} = (\mathbb{T} + 1)^{2}\mathbb{T}^{2}.
\end{equation}
\endproof

\begin{lem}\label{ban1}
$^{1}\G$ is the graph with two vertices and a pair of parallel edges, $e, f$, between them. 
Then the Grothendieck class is
\begin{equation}\label{ban1class}
\{\mathbb{V}_{^{1}\G}\} =  (\mathbb{T} + 1)^{2}\mathbb{T}^{2} 
+ (\mathbb{T} + 1)^{2}\mathbb{T}^3 + \mathbb{T}(\mathbb{T} + 1)(\mathbb{T}^{2} + \mathbb{T} + 1).
\end{equation}
\end{lem}

\proof
The \textbf{V}-polynomial of this graph is
\begin{equation}\label{Vban1}
V_{^{1}\G} = x_{c1+c2}(t_{e} + t_{f} + t_{e}t_{f}) + x_{c1}x_{c2}.
\end{equation}
If $x_{c1 + c2} = 0$, then $x_{c1} x_{c2} \neq 0$ and $t_{e}$ and $t_{f}$ are free, contributing a class 
\begin{equation}
\mathbb{L}^{2}(\mathbb{L} - 1)^{2} = (\mathbb{T} + 1)^{2}\mathbb{T}^{2}
\end{equation}
If $x_{c1 + c2} \neq 0$, then $x_{c1+x2}(t_{f} + 1)t_{e} \neq -x_{c1}x_{c2} - t_{f}$.
So if $t_{f} \neq -1$, this is equivalent to $$t_{e} \neq \frac{-x_{c1}x_{c2} - t_{f}}{x_{c1+x2}(t_{f} + 1)},$$ contributing a class 
\begin{equation}
\mathbb{L}^{2}(\mathbb{L} - 1)^{3} = (\mathbb{T} + 1)^{2}\mathbb{T}^{3}.
\end{equation}
If $t_{f} = -1$, then this is equivalent to $x_{c1}x_{c2} \neq -1$, contributing a class 
\begin{equation}
\mathbb{L}(\mathbb{L} - 1)((\mathbb{L} - 1)^{2} + \mathbb{L}) = \mathbb{T}(\mathbb{T} + 1)(\mathbb{T}^{2} + \mathbb{T} + 1).
\end{equation}
Therefore, in total, we obtain \eqref{ban1class}.
\endproof

We then obtain the following.

\begin{thm}\label{bananathm}
The Grothendieck class $\{\mathbb{V}_{^{m}\G}\}$ of the hypersurface complement for
the Potts model with magnetic field on the $m$-th banana graph is given by
\begin{equation}\label{bananas}
\{\mathbb{V}_{^{m}\G}\} = -(1+\mathbb{T})\mathbb{T}^{m+1}
+ \mathbb{T}(\mathbb{T} + 1)^{m+3}.
\end{equation}
\end{thm}

\proof Using Lemmata \ref{ban0} and \ref{ban1}, together with the generating function \eqref{gen}, we obtain
$$ \{\mathbb{V}_{^{m}\G}\} = ((\mathbb{T} + 1)((\mathbb{T} + 1)\mathbb{T}^{2} + (\mathbb{T} + 1)^{2}\mathbb{T}^{2}) - ( (\mathbb{T} + 1)^{2}\mathbb{T}^{2} + (\mathbb{T} + 1)^{2}\mathbb{T}^{3} + \mathbb{T}(\mathbb{T} + 1)(\mathbb{T}^{2} + \mathbb{T} + 1)))\mathbb{T}^{m} $$
$$ + (((\mathbb{T} + 1)^{2}\mathbb{T}^{2} + (\mathbb{T} + 1)^{2}\mathbb{T}^{3} + \mathbb{T}(\mathbb{T} + 1)(\mathbb{T}^{2} + \mathbb{T} + 1)) - \mathbb{T}((\mathbb{T} + 1)\mathbb{T}^{2} + (\mathbb{T} + 1)^{2}\mathbb{T}^{2}))(\mathbb{T} + 1)^{m}
$$
Simplifying, we obtain \eqref{bananas}.
\endproof

\smallskip

\begin{rem}\label{bananarem}{\rm
One should compare \eqref{bananas} with the case without magnetic field described in Example 6.6 of \cite{AluMa}.
The Grothendieck class of the hypersurface complement of the $m$-th banana graph, in the case of the Potts model
without magnetic field is given by $\bT^m +(\bT-1)(\bT+1)^{m+1}$. While the presence of the magnetic field does not
change the form of the recursion and the generating function, it does affect the individual terms, hence the resulting
expression as a function of $\bT$.
}\end{rem}

\medskip
\subsection{The Euler characteristic with compact support}
The Euler characteristic factors through the Grothendieck ring. In the example of the banana graphs,
we can then compute the Euler characteristic with compact support $\chi_{c}(\mathbb{V}_{^{m}\G}(\mathbb{R}))$ of 
the set of real zeros $\mathbb{V}_{^{m}\G}(\mathbb{R})$ of the partition function of the Potts model
with magnetic field.

\begin{prop} 
For the $m$-th banana graph $^{m}\G$, the Euler characteristic with compact support 
$\chi_{c}(\mathbb{V}_{^{m}\G}(\mathbb{R}))$ is given by
\begin{equation}\label{EulCh}
\chi_{c}(\mathbb{V}_{^{m}\G}(\mathbb{R})) = (-2)^{m+1}+ (-1)^{m}2
\end{equation}
\end{prop}

\proof This follows directly from \eqref{bananas} using the fact that $\chi_c(\bT)=\chi_c(\bG_m(\R))=-2$,
and the fact that $\chi_c$ defines a ring homomorphism $\chi_c: K_0(\cV_\R)\to \Z$.
\endproof

The expression for the Euler characteristic  with compact support is concise, reflecting the simplicity of
the banana graphs, but the rate of growth is exponential, reflecting the complexity of 
the \textbf{V}-polynomial.

\section{\textbf{V}-polynomial evaluation (computational complexity)}\label{CompuSec}

The \textbf{V}-polynomial grows exponentially with the size of the graph. Thus, rather than considering
the complexity of the problem of providing the entire \textbf{V}-polynomial of a graph, we focus on other
related computational complexity questions. In the literature on the complexity of various 
combinatorial polynomials of graphs that arise in relation to Potts models, problems of 
computing a specific coefficient of the polynomial and of evaluating the polynomial at a point
have been characterized, see \cite{NobWel}. Since for the \textbf{V}-polynomial all the coefficients
are equal to one, the relevant complexity problem remains that of evaluating the polynomial at a point. 
This problem is important because for the cases in which evaluating the polynomial at a point is tractable, 
determining whether a point is a phase transition is also tractable.

\smallskip

The specific problem is determining, given a graph $\G$ and the specification of a point where all 
variables not included are assumed to be zero, whether the value of the \textbf{V}-polynomial at 
that point is greater than a number $N$.

\medskip
\subsection{Potts model with magnetic field on the line graph and on polygons}

In \cite{NobWel}, computing the partition function for a restricted case of the \textbf{W}-polynomial 
was shown to be NP-hard for the star graph, so we investigate the complexity of more restricted trees. 
The first tree we consider is the 1-D Potts model or line graph, which is the tree with $n$ 
consecutive vertices, with every vertex but the first and the last having two neighbors each.

\smallskip

\begin{thm}\label{thmVline}
Evaluating the \textbf{V}-polynomial for the line graph is tractable
\end{thm}

\proof
We can see that the Fourtuin-Kastelyn representation \eqref{Vpoly} for the polynomial has 
exponentially many terms in the size of the line. We  label the vertices $1$ through $n$ from left to right, 
and label the edges $e_{1}$, $e_{2}$, ..., $e_{n-1}$ also from left to right. Then the 
Fourtuin-Kastelyn representation is
\begin{equation}\label{1Dgraph}
\sum\limits_{A \subseteq E} x_{c_{1}}x_{c_{2}} ... x_{c_{k(A)}}	\prod_{e \in A} t_{e}
\end{equation}
where $c_{i}$ is the sum of the weights of all the vertices in the $i$-th connected component of 
$A$, and $k(A)$ is the number of connected components. 

\smallskip

Thus, to every connected component that appears in a subset of edges $A$ we can associate a 
product of terms that appears in the sum $\chi_{i,j} = x_{c_{i,j}} \prod_{k=i}^{j} t_{e_{k}}$, where 
$c_{i,j}=\sum_{k=i}^{j} w_{k}$, with $w_{i}$ the weight of the $i$-th vertex. 
We can think of $\chi_{i,j}$ as corresponding to the connected component that starts 
at vertex $i$ and ends $j$ with the vertices also labeled from left to right. There are only $O(n^{2})$ 
connected components, so the values of these connected components can be computed and 
stored in a table in polynomial time. Now the sum can be rewritten as
\begin{equation}\label{1Dgraph2}
\sum\limits_{A \subseteq E}\prod_{k=1}^{k(A)} \chi_{i_{k},j_{k}}.
\end{equation}

\smallskip

Computing the sum naively still would require a prohibitive amount of time, since there are exponentially many terms. 
So we use the recursive structure of the line. Let $V_{i}(x,t)$ be the \textbf{V}-polynomial of the sub-line starting at vertex i. Every subset $A \subseteq E$ has a first connected component, namely the connected component containing $v_{1}$, which can be represented by $\chi_{1,i}$. Thus, we can write our sum for $V_{1}(x,t)$ as
\begin{equation}\label{1Dgraph3}
V_{1}(x,t) = \sum\limits_{i=1}^{n} \chi_{1,i}V_{i+1}(x,t).
\end{equation}
Now, we use dynamic programming to compute the sum in $O(n^{2})$ time. 
In round $j$, assume that we have the values of $V_{j+1}$ through $V_{n}$ precomputed. 
Then we can compute 
\begin{equation}\label{1Dgraph4}
V_{j}(x,t) = \sum\limits_{i=j}^{n} \chi_{j,i}V_{i+1}(x,t)
\end{equation}
in $n - j$ steps. Therefore, the number of steps it takes to compute rounds $1$
through $n$ is $$\sum_{i=1}^{n} i = O(n^{2}).$$
\endproof

\smallskip

As an easy consequence, we see that polygonal graphs
are also tractable.

\begin{cor}
Evaluating the \textbf{V}-polynomial on the polygonal graphs is tractable.
\end{cor} 

\proof Let $^{(n)}\G$ be the circle graph with $n$ edges. It is tractable since 
$^{(n)}\G\smallsetminus e$ is the line graph with $n-1$ edges, and $^{(n)}\G/e = ^{(n-1)}\G$. 
So $T(n) \leq T(n-1) + O(n^{2})$, where $T(n)$ is the number of steps to evaluate $^{(n)}\G$. Therefore 
\begin{equation}
T(n) \leq \sum_{i=1}^{n} O(n^{2}) \in O(n^{3})
\end{equation}
\endproof

\medskip
\subsection{Potts model with magnetic field on full binary trees}

The next case we consider is the full binary tree. In contrast to the previous cases, we have the following result.

\begin{thm}\label{bintreethm}
Evaluating the \textbf{V}-polynomial for the binary tree is NP-hard.
\end{thm}

\proof
We reduce from $\frac{1}{2}$-partition. The problem is, if we are given a finite subset $S \subset \mathbb{N}$, 
to determine whether we can partition $S$ into two parts with equal sums. Given a subset S of size $n$, let 
$M = \sum_{s \in S} s$. Then create a binary tree of height $\lceil \log(n)\rceil + 1$. Now number $S$ and set 
the vertex weights for the first $n$ leaf nodes to be $s_{i} \in S$. Set the weights of the rest of the leaf nodes 
to zero and set the vertex weights of all other nodes to $\lfloor \frac{3}{2}M \rfloor$. 
We then evaluate the polynomial at a point where we set $x_{s_{i}} = 1$, $t_{e} = 1$, $x_{m\lfloor \frac{3}{2}M \rfloor + \lfloor \frac{1}{2}M \rfloor} = 1$, for all $m$ such that $m \leq 2^{\lceil \log(n)\rceil}$, with all the other $x$ variables set to zero. 

The \textbf{V}-polynomial evaluates to something greater than 0 if and only if there exists a 
$\frac{1}{2}$-partition for $S$. To see this, observe the Fourtuin-Kastelyn representation of the polynomial 
\begin{equation}\label{bintreeVpoly}
\sum\limits_{A \subseteq E} x_{c_{1}}x_{c_{2}} ... x_{c_{k(A)}}	\prod_{e \in A}t_{e}
\end{equation}
Now, if any term is non-zero, then that means that, since $A$ has edges, for some $i$ we have $x_{c_{i}} = 1$, 
with $c_{i} = m\lfloor \frac{3}{2}M \rfloor + \lfloor \frac{1}{2}M \rfloor$. The only way that this could happen 
would be as a contribution of $\frac{1}{2}M$ from the leaf nodes. 

Conversely, suppose there exists a subset $B \subseteq S$, such that $\sum_{b \in B} b = \frac{1}{2}M$. 
Then look at the subset $A$ of $E$ in which we take all of the edges above the nodes labelled with elements in $B$, as well as all of the other edges in the tree. The $x_{c_{i}}$ variables in that term will correspond either to the unique non-singleton connected component defined by the edge subset $B$, or to some $s_{i} \in S$.
In the first case, there are an odd number of nodes with vertex weight $\frac{3}{2}M$ and leaf nodes with 
vertex weights that contribute another $\frac{1}{2}M$. In the second case, all the variables in the term are 
equal to one. Therefore, we have
\begin{equation}\label{bintreeVpoly2}
\sum\limits_{A \subseteq E} x_{c_{1}}x_{c_{2}} ... x_{c_{k(A)}}	\prod_{e \in A}t_{e} \geq 1.
\end{equation}
\endproof

\medskip
\subsection{Trees limiting to the line graph}

The method used in Theorem \ref{bintreethm} is sufficiently general that we can extend it to a 
class of trees that limits to the line graph.

\begin{defn}\label{1ntree}
A tree is a $1 + \frac{1}{n}$-ary tree, if $\frac{1}{n}$ of its non-leaf nodes have two children, with the rest having one child. 
\end{defn}

For instance, a $1 + 1$-ary tree would be a binary tree. 
Note that, in general, there are many non-isomorphic $1 + \frac{1}{n}$-ary trees of the same size. 
However, if two $1 + \frac{1}{n}$-ary trees are the same size, then they have the same number of leaves. 
We can see this because the number of leaves is equal to the number of nodes with $2$ 
descendents plus $1$, by induction. In particular, for all $n$, the number of leaves increases 
linearly with the size of the $1 + \frac{1}{n}$-ary tree. The following statement takes into account
this ambiguity in the non-isomorphic trees.

\begin{thm}
For all $n \in \mathbb{N}$ evaluating the \textbf{V}-polynomial on a family of $1 + \frac{1}{n}$-ary trees 
with elements of every possible size is NP hard.
\end{thm}

\proof
Suppose given an instance of $\frac{1}{2}$-partition of size $|S| = m$. Place the nodes on the leaves 
of one of the $1 + \frac{1}{n}$-ary trees, $T$, with a minimal number of nodes, while still having more than 
$m$ leaves.  Set $x_{s_{i}} = 1$, $t_{e} = 1$, and  $x_{m\lfloor \frac{3}{2}M \rfloor + \lfloor \frac{1}{2}M \rfloor} = 1$, 
where $m$ is less than the number of nodes of $T$. Then, by same reasoning as above, the polynomial 
evaluates to something greater than zero if and only if there exists a $\frac{1}{2}$-partition.
\endproof

This shows that the binary tree is NP-hard not because of its completeness, but because it has a 
constant proportion of branching.

\bigskip

\noindent {\bf Acknowledgment} The first author was supported 
by a Summer Undergraduate Research Fellowship at Caltech. The second
author is supported by NSF grants DMS-1007207, DMS-1201512, PHY-1205440. 
The second author thanks Paolo Aluffi and Spencer Bloch for useful conversations.

\end{document}